# Characterization of a commercially available large area, high detection efficiency single-photon avalanche diode

Mario Stipčević,[1,2,3*] Daqing Wang,[4] and Rupert Ursin[4]

[1]*University of California Santa Barbara, Dept. of Physics, Santa Barbara, CA 93106-9530, USA*
[2]*Duke University, Dept. of Physics, Durham, NC 27708, USA*
[3]*On leave from Ruđer Bošković Institute, Dept. of Experimental Physics, PP 180, 10002 Zagreb, Croatia*
[4]*Institute for Quantum Optics and Quantum Information (IQOQI), Austrian Academy of Sciences, Boltzmanngasse 3, A-1090 Vienna, Austria*
[*]*Mario.Stipcevic@irb.hr*

*Abstract* - We characterize a new commercial, back-illuminated reach-through silicon single-photon avalanche photo diode (SPAD) SAP500 (Laser Components. Inc.), operated in Geiger-mode for purpose of photon counting. We show that for this sensor a significant interplay exists between dark counts, detection efficiency, afterpulsing, excess voltage and operating temperature, sometimes requiring a careful optimization tailored for a specific application. We find that a large flat plateau of sensitive area of about 0.5 mm in diameter, a peak quantum efficiency of 73% at 560 nm and timing precision down to 150 ps FWHM are the main distinguishing characteristics of this SPAD.

*Index Terms—* **Photodiodes; Photodetectors; Quantum communication; Quantum detectors.**

## I. INTRODUCTION

Over the past two decades single pixel solid silicon avalanche photodiodes sensitive in visible and near-infrared (NIR) range have found a wide range of applications, both in linear (sub-Geiger) and photon counting (Geiger) mode of operation. Because of their small size, high gain, excellent timing, high immunity to magnetic fields and electromagnetic induction pickup, they replaced single-channel photomultipliers in almost all applications except those requiring large detection area beyond one millimeter, which is required in applications such as large area Cherenkov detectors [1] or free-space communication with direct detection [2]. Specifically, new detectors of this type are very interesting in research fields like quantum entanglement [3], quantum cryptography [4] and optical quantum computing.

There are two distinct types of avalanche (silicon) photodiodes: those capable of operation only in linear (sub-Geiger) regime (usually named just APD) and those capable (also) of operation above the breakdown voltage (commonly known as SPAD). In a SPAD biased above breakdown voltage a single photon may trigger a self sustained avalanche current that may last for a long time (tens of microseconds or longer) before self-ceasing. In order to restore SPAD's ability to detect a new photon it is necessary to quench the avalanche, similar to quenching of a well known Geiger-Muller gas tube where quenching is usually done by a passive electronic circuit. However, both to exploit and to explore all advantages and characteristics of a SPAD, a faster quenching mechanism based on active electronics components is necessary.

In our previous work [5] we have tested particularly promising pre-production samples of a SPAD made by Laser Components whose intended commercial name was SAP500. However, the commercial version of it was not available for almost about a year later and apparently some changes have been made that lead to improved flatness of the detection efficiency over the sensitive area, improved timing performance and enhanced response in near infra red (NIR) part of the spectrum due to the better anti-reflective coating of the chip. On the other hand, dark counts level has risen several times. We also noted that the short-circuit avalanche current of the commercial SAP500 is about 20 mA whereas of pre-production samples it was about 3 mA.

In the present study we have investigated the new, commercial SAP500 diode in TO-8 case which incorporates a two-stage thermoelectric cooler, a 1.5 kΩ NTC resistive temperature sensor and an anti reflective coated glass window. Using a temperature controller (Wavelength Electronics, WTC3243) we were able to set the operating temperature point of the SPAD chip between -25°C and +20°C. Using SAP500 as a sensor and a home-made active avalanche quenching (AAQ) circuit similar to one described in [6] we have built a single-photon detector used in all measurements except that of the spatial profile of detection efficiency (Sect. 7) for which we used a simple passive quenching circuit. The AAQ circuit features small dead time (26 ns) and fast quenching voltage

We thank Xiaosong Ma, Sebastian Kropatschek and Thomas Herbst for assistance in the lab. This work was made possible by grants from the European Space Agency (contract 4000104180/11/NL/AF), the Austrian Science Foundation (FWF) under projects SFB F4008 and CoQuS, and the FFG for the QTS project (no. 828316) within the ASAP 7 program. We also acknowledge support by the European Commission, grant Q-ESSENCE (no. 248095) and the John Templeton Foundation. MS was supported by Croatian ministry of science grant no. 098-0352851-2873.

M. Stipčević (corresponding author) is with Rudjer Boskovic Institute, Bijenicka 54, 10002 Zagreb, Croatia (e-mail: mario.stipcevic@irb.hr).

R. Ursin and D. Wang are with Institute for Quantum Optics and Quantum Information (IQOQI), Austrian Academy of Sciences, Boltzmanngasse 3, A-1090 Vienna, Austria.

transitions (a few ns) so that it has a negligible influence on the measured characteristics of the SPAD which have been made at detection frequencies of up to 100 kcps.

## II. SAP500 STRUCTURE

The structure of SAP500 shown in Fig. 1 is a typical reach-through, illuminated from the back side [7]. A special characteristic of this SPAD is that the conversion region, situated between p+ layer and p+ region, is rather thin – only about 25 μm – but the junction side of the SPAD is covered with a metal layer that acts as a mirror and effectively duplicates the length of the available conversion path thus improving the quantum efficiency in red and NIR part of the spectrum. On top of that a broad-band anti-reflective (AR) coating of the entrance surface minimizes reflectance for the incoming photons. The short conversion region improves timing precision of incoming photons (jitter) by ensuring smaller drift time dispersion and higher carrier velocity. Another benefit of short conversion region is possibility of operation at a low breakdown voltage which is typically 125 V at 22 °C, while other SPADs with similar spectral QE typically operate between 250 V and 500 V. Low operating voltage is favorable for smaller heat dissipation of the SPAD at high counting rates and simpler avalanche quenching electronics.

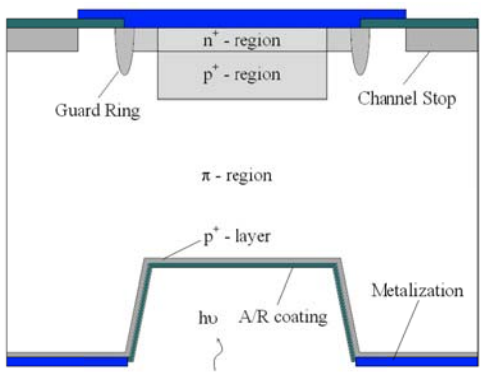

Figure 1. Cross section of the SAP500. It is a back-illuminated reach-through structure. The photon conversion region is situated between $p^+$ layer and $p^+$ region and is only 25 μm thick. bottom and inactive part of the top are covered by metalized layers whose purpose is to enhance conversion of photons.

## III. DEAD TIME

The dead-time of a photon detector is defined as the time span between the impact of a photon and the time when the detector is ready for sensing the next photon. It is mainly determined by the electronics employed. Our avalanche quenching circuit is capable of reliably quenching the SAP500 diode at an overvoltage of up to 15V and has a dead time of about 26 ns Fig. 2. A short dead time is generally favorable for high photon detection rates but it can lead to a larger afterpulsing, as will be discussed later.

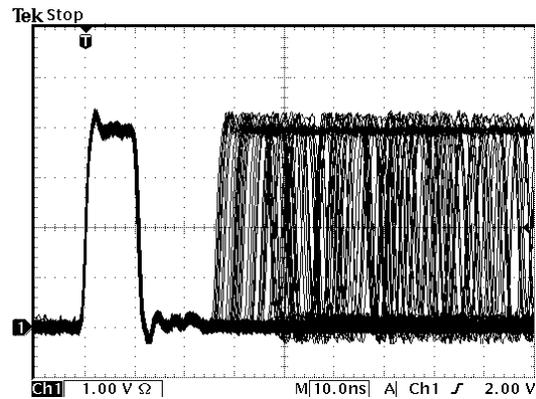

Figure 2. Oscillogram of the cumulative output from the photon detector: a dead time of ≈26ns is clearly visible between the leading edges of two successive output pulses from the detector.

## IV. BREAKDOWN VOLTAGE

Breakdown voltage of a SPAD is the lowest voltage at which a single free carrier can trigger a self-sustained avalanche. When the bias voltage of an APD is raised one expects appearance of dark counts just as the bias exceeds the breakdown level. However the dark count rate vanishes because the detection efficiency near breakdown tends to zero. To facilitate the measurement, we shine an attenuated beam of light from a 676 nm CW operated light-emitting diode (LED) yielding a flux of $(20.0\pm0.1)\times10^3$ photons per second on the SPAD. Under that condition, when the operating voltage is scanned around the breakdown level emerges as a sharp ON/OFF avalanche characteristic, allowing to determine the breakdown voltage with a resolution of about 10-20 mV. We find that the breakdown voltage is a linear function of temperature with a coefficient of only $(0.273\pm0.024)$ V/K, see Fig. 3, which is quite small for a reach-through SPAD. Low temperature coefficient facilitates achieving a good temperature stability of the operating point.

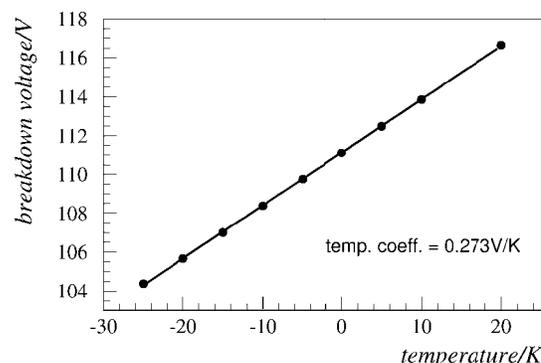

Figure 3. Breakdown voltage is a linear function of temperature characterized by the temperature coefficient which we measured to be 0.273 V/K for the particular sample.

## V. DARK COUNTS

Dark counts are avalanches that appear even when SPAD is kept in total darkness (sometimes also referred to as intrinsic dark counts). They happen randomly in time. In applications where photon arrival time is unknown (and therefore signal cannot be extracted by coincidence techniques [8]), dark counts present an irreducible physical background noise which is indistinguishable from the useful signal and thus limit both the lowest detectable photon rate *and* the dynamic range of detectable signal. During this measurement the SPAD is kept in complete darkness. Our setup allows to independently set the operating voltage and temperature of the SPAD. Figure 4 shows measured dark count rate as a function of the overvoltage while SPAD was kept at -20 °C. The stars are the measured points and the curve is least squares parabola fit.

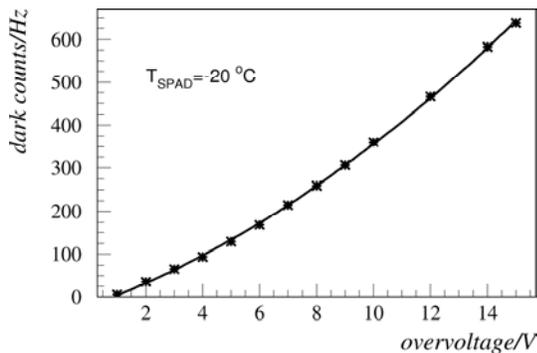

Figure 4. Dark counts rate as a function of the overvoltage at SPAD temperature of -20°C.

We notice that the experimental points are neatly aligned on a parabola which, of course, intersects point (0,0) (zero dark counts at zero overvoltage, as mentioned above). As of yet we have no precise quantitative model that would explain this precise parabolic behavior but a similar dependence has already been observed for an InGaAs/InP APD [9]. In a silicon SPAD operated near the room temperature the primary dark count rate (DCR) has two contributions: thermal and filed-assisted (eg. tunneling) generation of carriers. On top of that we have afterpulses as a secondary mechanism of dark counts but their contribution is typically limited to a few percent. If thermal excitation would be the dominant mechanism, then the concentration of carriers at the fixed temperature would be constant. In that case the DCR curve would qualitatively follow the detection efficiency curve which is characterized by a sharp rise at low overvoltages and saturation (decreasing derivative) at high overvoltages, as shown in Fig. 7. However our measurement in Fig 4. show that the DCR rises parabolically (increasing derivative) indicating that the field-assisted component plays a dominant role.

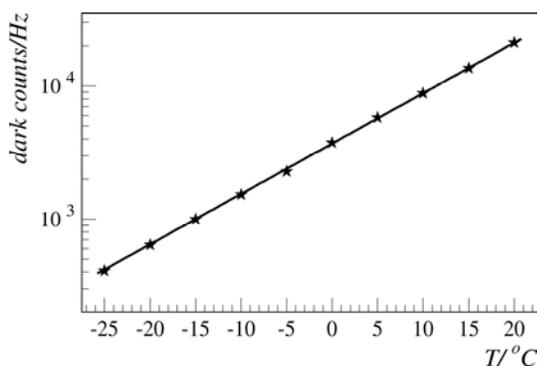

Figure 5. Dark counts rate as a function of the temperature at overvoltage of 15 V. Straight line is an exponential least squares fit to the data points.

Free carriers in the depleted multiplication region of the SPAD, which can cause dark counts, show exponential dependence of their density on the temperature, as indeed confirmed by measurements shown in Fig. 5. Data points have been taken at a constant overvoltage of 15 V. We see that cooling the SAP500 from the room temperature of +20 °C to -25 °C reduces the dark counts rate by a factor of ≈52. Note that lowering the temperature helps eliminate dark counts but at the same time it enlarges afterpulsing probability. Because both effects are non negligible in SAP500, choice of the optimal operating temperature may be an important concern for some applications.

## VI. PHOTON DETECTION EFFICIENCY

Photon detection efficiency (PDE) is one of the most important parameters of any single-photon detector. It is defined as the probability that detector generates a digital output pulse corresponding to an incoming single photon. It is often mistaken for quantum efficiency (QE), which is a probability of conversion of a photon into a charge carrier(s) in the detector's sensitive volume. While QE is predominantly a property of material itself and depends very little on operating conditions, detection efficiency is a product of QE and probability that a converted photon will generate an avalanche and depends dramatically on the operating overvoltage, $V_{OVER}$. Clearly, just below the breakdown voltage ($V_{OVER} = 0$), the detection probability is equal to zero because no self-sustained avalanche can exist. In another words, even though photons are converted into carriers with efficiency equal to QE (which can be as high as 95% at 560 nm [10]), none of the carriers will trigger a self-sustained avalanche, which means that the detection efficiency is zero.

The detection efficiency is defined as the average frequency of photon detections divided by the average frequency of incoming photons. The average frequency of incoming photons has been estimated by means of a single-photon detector of known detection efficiency, and is kept constant during the measurement. The crucial point of this measurement is to count only detections of incoming photons. In order to isolate real photon detections (which contribute to the PDE) and suppress afterpulses and noise (which do not contribute to the PDE), we use a setup with a picoseconds pulsed laser shown in Fig. 6 which counts only pulses that are in a tight coincidence window (5.0 ns) with the incoming photons. Figure 7 shows PDE as a function of overvoltage for photon wavelength of 676 nm. It reaches 66% at 15 V overvoltage which corresponds to a maximum value of 73% at 560 nm. The efficiency at 676 nm was measured relative to the efficiency of a detector previously calibrated by the source of entangled photons [5], while the efficiency at 560 nm was obtained by scaling the value at 657 nm according to the quantum efficiency plot in the datasheet [10]. An interesting characteristic of the SAP500 illustrated in Fig. 7 is a very fast initial rise of PDE: at about 2 V of overvoltage PDE already reaches a half of the value at 15 V where it is close to saturation.

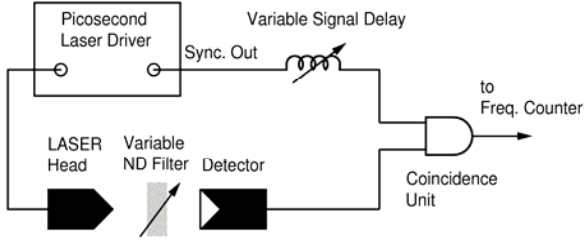

Figure 6. Setup for measuring relative detection efficiency. Attenuated pulses from picoseconds laser head are detected by the detector in a tight coincidence window (5.0ns) with the laser sync. signal.

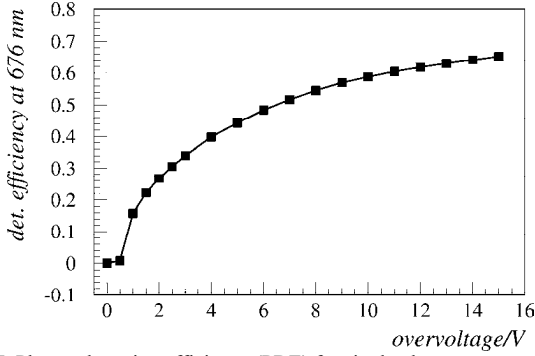

Figure 7. Photon detection efficiency (PDE) for single photons at a wavelength of 676 nm reaches 66% at 15V overvoltage. At an overvoltage passed 0.5V PDE starts to rise abruptly and then slows down and goes into asymptotic saturation at high overvoltages.

## VII. SPATIAL PDE PROFILE

We measured relative PDE versus the impact of a single photon at different positions on the active area, using a probe beam at 810 nm forming a 20 μm beam waist and performing an x-y scan relative to the detector position. The photon counting rate at each position was measured using an attenuated CW laser and a fiber-beam-splitter (FBS) for normalization. The SAP500 SPAD was operated in a passive quenching circuit during this measurement. The sensitive area was placed in the focus of one fiber splitter output and the other output was connected to a reference detector. Figure 8 shows the relative detection efficiency calculated as the ratio between count rates of the two detectors normalized to its peak value. The ratio is therefore insensitive to eventual intensity instability of the laser. We see that the relative PDE exhibits a flat top of about 500 μm in diameter and a sharp drop at the edges This feature is particularly interesting in free-space experiments such as [2], where the impact position is randomly spread over the active area of the detector due to atmospheric turbulences.

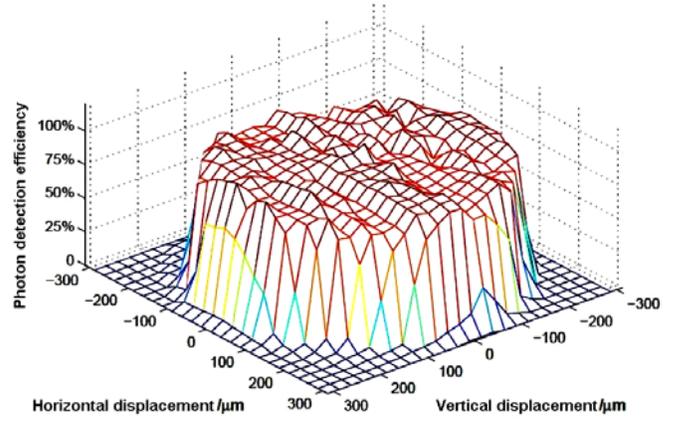

Figure 8. 2D spatial relative photon detection efficiency profile for a passively quenched SAP 500 SPAD operated at $V_{OVER}$ = 15 V.

## VIII. AFTERPULSING

During an avalanche, a tiny fraction of carriers may become trapped in metastable energy states in the bandgap where they can survive the quenching procedure, resurface at a later time and cause another avalanche called "afterpulse". A probability $p_a$ that an avalanche will be followed by an afterpulse is called "afterpulsing probability". Following [11] we model the avalanche probability density function $A(t)$ as:

$$A(t) = \eta(t - t_{dead})\left[R_{dark} + R_{phot} + P_c \frac{N_{ft}}{\tau_a} e^{-t/\tau_a}\right] \qquad (1)$$

where time $t$ is counted from the last avalanche, $R_{dark}$ is the dark count rate from the random thermal and tunneling processes, $R_{phot}$ is the rate of detected real photons, $P_c$ is a probability that a carrier will start an avalanche, $N_{ft}$ is number of filled traps, $\tau_a$ is the lifetime of trapped carriers and $t_{dead}$ is the dead time of the detector. In this model we assume only one dominant trapping mechanism which is the case for SAP500. Afterpulsing is modeled by the last, exponential term.

We measured the afterpulsing probability as a fraction of afterpulses in total pulses, in the limit of low detection frequency (about $20\times10^3$ counts per second). The setup for measuring afterpulses shown in Fig. 9 features a picosecond pulsed laser driven at 1 MHz. Detection of a real photon is identified by means of a tight coincidence time-window with laser sync pulse and corresponding electrical pulse is sent to the Output B. Each real photon event triggers a 500 ns long interval (starting at the end of the dead time) during which any subsequent events (afterpulses and dark counts) are sent to the Output A. The time intervals between events at Output B and Output A are measured by means of a time-to-amplitude converter (Ortec, 567) and sent to a PC computer. The setup was used in two ways: 1) for measuring time intervals between real photon detections and 2) for determining the afterpulsing probability.

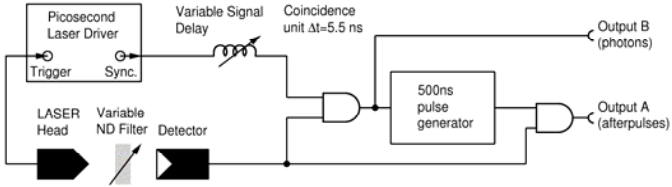

Figure 9. Setup for measuring afterpulses. Pulses of light from the picoseconds laser (Picoquant, 39 ps FWHM, 676 nm) are emitted at the rate of 250 kHz and are attenuated to 0.1 photon/pulse level by means of the variable neutral density filter. The setup separates detections of real photons from aftrepulses and sends each to its own output port.

A distribution of intervals obtained at 15 V overvoltage and SPAD temperature of -5 °C is shown in Fig. 10. According to the theoretical model (Eq. 1), releasing of trapped carrier is a Poissonian process and lifetime of the trapped carriers ($\tau_a$) decreases with temperature. The long right-hand tail present in the distribution of afterpulses, shown in Fig. 10, indicates that there are more than one carrier trapping mechanisms but one with the lowest lifetime is dominant. We assess its lifetime by an exponential fit to the leftmost part of the distribution and find that $\tau_a$ spans from 13.4 ns at +20 °C to 23.1 ns at -25 °C.

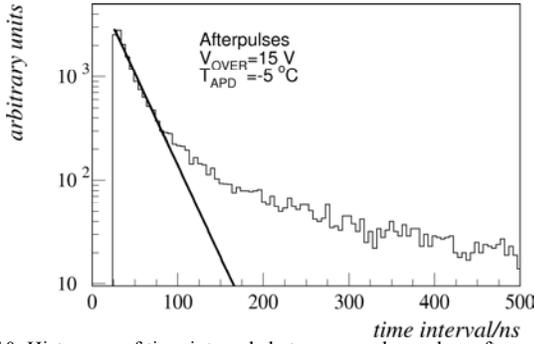

Figure 10. Histogram of time intervals between a pulse and an afterpulse. The background floor belonging to dark counts has been subtracted. Shown is also an exponential fit of the afterpulsing intervals which yields the fitted value of the afterpulsing lifetime, $\tau_a$.

We model the temperature dependence of the trap lifetime with Arrhenius' equation:

$$\tau_a = \frac{1}{\sigma v N} e^{E_a/kt} \qquad (2)$$

where further assuming constant trap cross section $\sigma$, density of states $N(T) \propto T^{3/2}$ and average velocity of carriers $v \propto T^{1/2}$ [12]. We expect a linear dependence between $\ln(T^2\tau_a)$ and $1/T$. Figure 11 shows a plot of measured points obtained by the procedure described above along with a least squares linear fit. It can be shown that the derivative of this (linear) curve is equal to $E_a/k$ which allows us to estimate the activation energy $E_a \approx 0.028$ eV.

With our setup shown in Fig. 9, afterpulsing probability ($p_a$) can be evaluated in a very simple manner by measuring the ratio $r_{AB}$ between pulse rates at the Output A and the Output B.

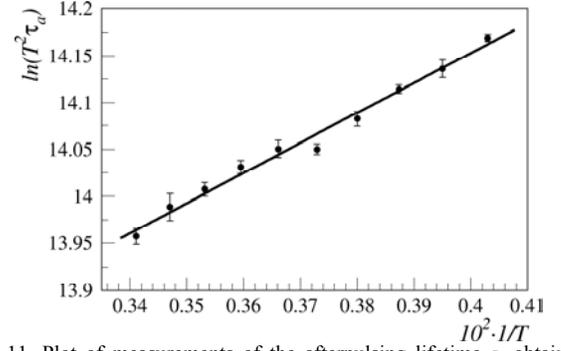

Figure 11. Plot of measurements of the afterpulsing lifetime $\tau_a$ obtained at various temperatures $T$ in the range from 248.2K … 243.2K (-25 °C … +20 °C). Shown are 1 sigma error bars.

We measured the ratio by a frequency ratio meter (Hameg HM8123). In absence of dark counts, afterpulsing probability would be exactly equal to the ratio $r_{AB}$, however because Output A also contains dark counts, a small correction should be applied as follows:

$$p_a = r_{AB} - \Delta T f_{DC} \qquad (3)$$

where $\Delta T$=500 ns is the length of the time interval during which afterpulses (and dark events) are counted and $f_{DC}$ is the mean frequency of dark counts at a given temperature and overvoltage. The counting period is chosen long enough (~30 times the lifetime) to safely account for all primary afterpulses and most afterpulses of afterpulses, yet the noise correction term is small (0.02% - 1%). Expression (3) neglects yet higher order effects such as afterpulses of dark counts.

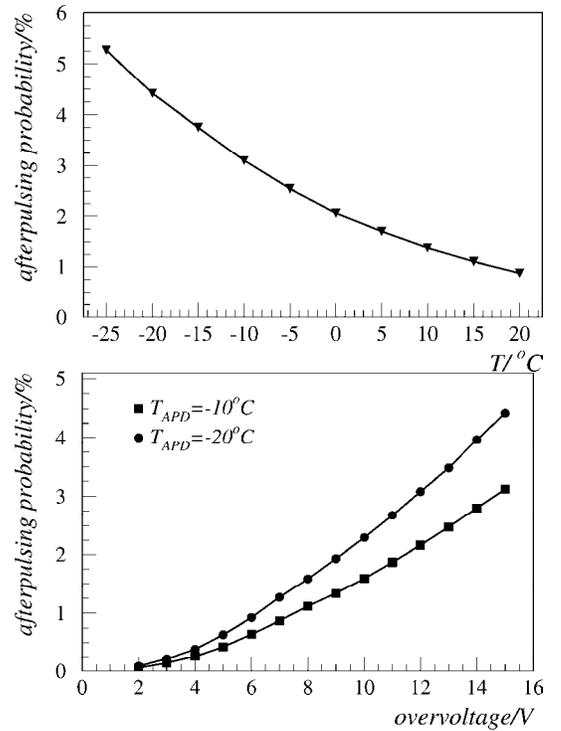

Figure 12. Afterpulsing probabilty within 500 ns after the dead time (26 ns) for SAP500: as a function of temperature at 15 V overvoltage (up); as a function of overvoltage at -10 °C and -20 °C (down). Only "visible" afterpulses, namely those happening after the dead time are counted.

From this, we find that for SAP500 the afterpulsing probability is a function of temperature, as shown Fig. 12 up. Namely because the afterpulsing probability density function (e.g. Fig. 10) sharply decreases in time, a significant fraction of afterpulses is "lost" in the dead time of the detector and only those surviving the dead time are "visible". As temperature of the SPAD rises, trapping lifetime decreases and the fraction visible afterpulses decreases.

We also measured the afterpulsing probability as a function of overvoltage, as shown in Fig. 12 down. The afterpulsing rises at high count rates as a function of overvoltage. The afterpulsing of SAP500 is comparably strong. It can be limited by a longer dead time or a higher operating temperature but none of these measures are without caveats. Therefore, for certain applications an optimized trade-off may be required. In quantum cryptography, for example, longer dead time directly reduces the secret key rate while afterpulsing increases bit error rate (BER).

## IX. PHOTON ARRIVAL TIME JITTER

One of the most fascinating characteristics of SAP500, when compared to other known SPADs with similar quantum efficiency in the visible and near infrared range, is its superior timing performance. In this measurement we send weak light pulses (about 0.01 photon per pulse) from the picoseconds laser and measure the time between photon emission and its detection by the detector. SAP500 shows small jitter already at a low overvoltage and it goes down to 150 ps full width half maximum (FWHM) at 15 V as shown in Fig. 13.

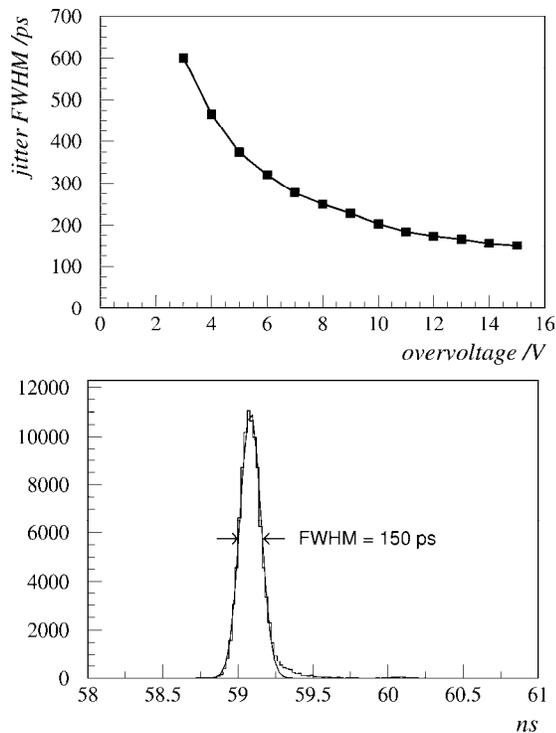

Figure 13. Jitter FWHM as a function of the overvoltage at -10 °C (up); histogram of time differences between photon arrival times and corresponding avalanches as detected by our AAQ circuit (down). The light at 676 nm was focused to spot size of 50 μm in the center of the APD.

The arrival time distributions are nearly Gaussian at any measured overvoltage. In the measurement the light was focused onto a spot of diameter of 50 μm FWHM in the center of the SPAD. We noted that the jitter does not change by tighter focusing but worsens if the spot size is larger than 50 μm. For example, for a light beam spread evenly over the whole sensitive area, the jitter at 15 V rises to 405 ps FWHM and exhibits a long right-hand tail as well as shift of the mean value towards longer times. This could be explained by uneven strength of the drift electric field over the conversion volume (π-region in Fig. 1). The shift and the tail indicate existence of regions of a slower carrier drift, that is with a lower electric field, situated at a larger radii from the center of the SPAD.

The radial dependence of timing is in contrast with the flatness of the detection efficiency over the whole nominal diameter of the SPAD (Fig. 8). This is because the avalanche multiplication region, delimited by the $n^+$ and $p^+$ layers (Fig.1) where a drifted carrier has a chance to trigger a self-sustained avalanche, is very thin and the field strength there is constant over the whole area.

## X. CONCLUSION

We have characterized a VIS-NIR commercial SPAD (Laser Components, SAP500) by means of a home-made electronics circuits. We studied breakdown voltage, dark counts, detection efficiency, 2-dimensional spatial detection efficiency profile, afterpulsing and timing jitter. The maximum overvoltage at which SPAD was operated was 15 V due to the fact that the important detection parameters came into saturation by that point while count rate and afterpulsing deteriorated substantially. For example, at -20°C going from 12 V to 15 V overvoltage the detection efficiency improved by mere 3%, jitter improved by 11% while both DCR increased by 37% an afterpulsing probability by 44%. This shows also, as expected, that some of the parameters can be improved by appropriate choosing of operating bias and temperature of the SPAD, but some of the improvements are mutually exclusive. Different optimizations may be preferred for specific applications. In a tentative "general purpose setting" (SPAD temperature -10 °C, overvoltage 15 V) we obtained detection efficiency of 66% at 676 nm (peak 73% at 540 nm), timing jitter of 150 ps FWHM, dark counts of 1.5 kHz and visible afterpulsing probability of 3.2%, which combined with a large flat response area of 0.5 mm diameter is a potentially very interesting performance for quantum communication and quantum information research and applications.